# Molecular biaxiality determines the helical structure - Infrared measurements of the molecular order in the $N_{TB}$ phase of difluoro terphenyl dimers

Katarzyna Merkel[1], Barbara Loska[1], Chris Welch[2], Georg H. Mehl[2], Antoni Kocot[1]

[1]Institute of Materials Engineering, University of Silesia, 75 Pułku Piechoty 1A, 41-500 Chorzów, Poland
[2]Department of Chemistry, University of Hull, Hull HU6 7RX, UK

**Abstract:** IR polarized spectroscopy was employed, to obtain the three components of IR absorbance for a series DTC5Cn (n=5,7,9,11) of bent-shaped dimers containing double fluorinated terphenyl core. Then, they were used to calculate both uniaxial (*S*) and biaxial (D) order parameters, for various molecular groups of dimers. The molecule bend was estimated due to observed difference between *S* parameters for the terphenyl core and central hydrocarbon linker. It is found to be significant in N phase despite almost no macroscopic director bend. The temperature dependence of the order parameter, *S*, distinctly reverses its monotonic trend of increase to decrease at the transition temperature, $T_N$-$N_{TB}$ to the $N_{TB}$ phase as result of the director tilt in the $N_{TB}$ phase. The biaxial order parameter, *D*, is negligible in the N phase, and then starts increasing on entering the $N_{TB}$ temperature, similarly to $\sin^2\theta$. The local director curvature is found to be controlled by the molecular biaxiality parameter.

**Introduction:**

An important leitmotif in the science of soft and liquid crystal materials is the understanding of the mutual relations between the shape of a molecule and macroscopic self-organization and the search for new possibilities of ordering based on the study of molecules of various shapes [1-8]. A fairly recent, and quite remarkable, manifestation of spontaneous chirality in the fluid of achiral molecules is the nematic twist-bend (TB) $N_{TB}$ liquid crystal phase, which is formed by long-range 1D ordered molecules following the heliconical precession of the director [9-15]. The structure is based mainly on the twist of the biaxial molecular order around the molecular long axes, quantitatively related to the curvature of the bend of the molecule, despite almost no macroscopic director bend. In order to elucidate the structure of the $N_{TB}$ phase, several models based on rigid molecules have been used. These were interpreted on the basis of nanophase segregation of the flexible central alkyl linkage/linker, terminal chains and rigid molecular subcomponents (cores): molecular ends find entropic freedom by associating with the flexible cores. This, together with the X-ray observation of the half-molecular length periodicity along $N_{TB}$ helix, lead to a proposal of a model of self-assembly of half molecule-long segments into duplex helical tiled chains of molecules as the basic structural element of the $N_{TB}$ phase [6,7,16]. Molecular arrangement, proposed for the CB7CB [15,17-19] and for series of DTC5Cn [20-23] appears to offer a useful benchmark for relating the molecular structure and macroscopic behavior of the $N_{TB}$ phases. Recently, the twist-bend nematic phase transitions in a series of DTC5Cn dimers with increasing spacer length were experimentally observed, mainly by a combination of experimental techniques such as: polarized microscopy (POM) [24-26], dynamic light scattering (DLS) [27], dynamic calorimetry and X-ray diffraction (XRD, SAXS, WAXS, GIXRD) [22,23,26], Resonant Soft X-Ray Scattering [26] and freeze-fracture transmission electron microscopy (FFTEM) [12]. Additionally, the arrangement of the $N_{TB}$ phase, the molecular dynamics, and order parameters were also studied for DTC5Cn dimers by electro-optical [24], dielectric [25] and nuclear magnetic resonance studies (NMR) [20,28].

In this paper we employ infrared spectroscopy to study the orientational order of the DTC5Cn terphenyl dimer series (n=5-11), for a number of molecular segments: the terphenyl core, central alkyl linker and the tail ends/terminal chains. By comparing the order parameters of the terphenyl core and the central alkyl linker, we examined the bend of the molecule in the nematic phase. In the $N_{TB}$ phase the order parameter reverses its trend due to the helical tilt of the director. Using the ratio of the $N_{TB}$ order parameter and the extrapolated trend from the nematic phase, the tilt angle for the different dimer segments can be calculated can also be useful to evaluate the local director bending.

**Experimental**

The chemical structure of 2´,3´-difluoro-4,4´´-dipentyl-p-terphenyl –n- alkanes (DTC5Cn, where n = 5, 7, 9, 11) are given in Figure 1. Details of synthesis and surface properties for these materials have been reported recently [26, ESI]. The sample for the IR study was aligned in between the two optically polished ZnSe windows. In order to obtain the homogeneous orientation of molecules, windows were spin coated with a SE-130 commercial polymer aligning (Nissan Chemical Industries, Ltd). The cells were assembled with parallel arrangement of the rubbing direction. In order to obtain the homeotropic alignment of samples we used a commercial solution of the AL 60702 polymer (JSR Korea). Mylar foil was used as a spacer and thickness of cells fabricated was determined to be in the range from 5.1 – 5.6 µm, by the measurements on the interference fringes using a spectrometer interfaced with a PC (Avaspec-2048). The sample was capillary filled by heating the empty cell in the nematic phase, five degrees below the transition to the isotropic phase.

The quality of the alignment has been tested using polarizing microscopy. The texture of the sample was monitored using a polarizing microscope that was used for identifying the phase prior to its investigation by polarized IR spectroscopy.

The infrared spectra are recorded using an Agilent Cary 670 FTIR spectrometer with a resolution of 1 cm$^{-1}$ and these spectra are averaged over 32 scans. An IR-KRS5 grid polarizer is used to polarize the IR beam. The polarized IR spectra are measured as a function of the polarizer rotation angle. Measurements were performed on slow cooling and heating at the rate of 0.1 K/min. Temperature of the sample was stabilized using PID temperature controller within ±2 mK.

The molecular structures of compounds were optimized and corresponding IR wavenumbers were calculated using the Gaussian 09 program, version E.01 [29]. Density functional theory (DFT) with Becke's hybrid three-parameter, Lee-Yang-Parr correlation functional, B3LYP and the diffusion and polarized basis set (6-311++ G**) have been used [30]. The possible structures of the rigid core of the each arm of the dimer can be classified by the dihedral angles ($\varphi_1$ and $\varphi_2$, see Fig. 1a) around the inter-ring in the terphenyl. The two most energy-favorable conformers of the terphenyl were established: helical ($\varphi_1$=-43.5, $\varphi_2$=-43.2) and twisted ($\varphi_1$=-43.2, $\varphi_2$=43.2) for which theoretical frequencies were calculated. All of the observed bands in the $N_{TB}$ phase been explained by the coexistence of the helical ($D_2$) and the twisted ($C_{2h}$) conformers. The supplementary material presents the calculated FTIR spectrum for one of the conformers (Fig. 1s) and the experimental spectrum in the $N_{TB}$ phase on the example of the DTC5C5 dimer (Fig. 2s). Some of the vibrational modes were found to be the most promising features for studying the orientational ordering of DTC5Cn molecules.

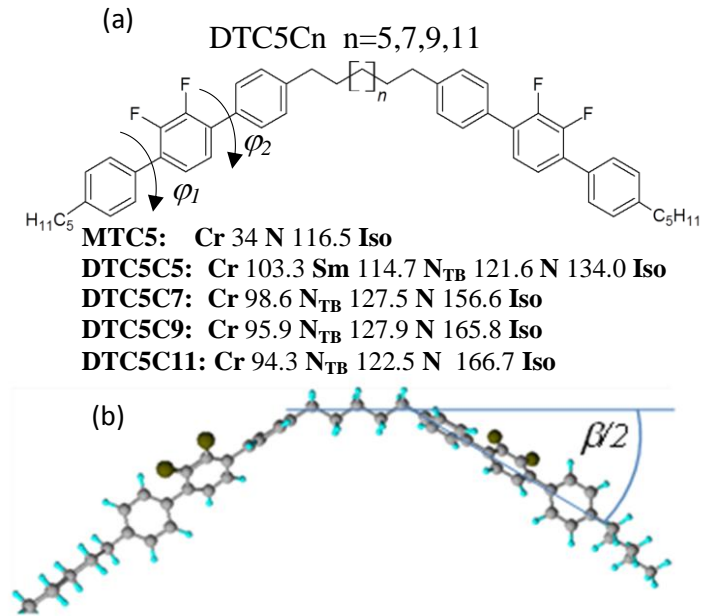

Fig.1. 2´,3´-difluoro-4,4´´-dipentyl-p-terphenyl dimers (DTC5Cn) (n=5,7,9,11). (a) Transition temperature and molecular structure of DTC5Cn. (b) Simulated structure of the helical conformer of a DTC5C5 molecule. $\beta_2$ – angle the core makes with the bow string axis of the dimer.

**The order parameters in terms of the components of the IR absorbance**

For an anisotropic system the use of vibration spectroscopy can provide information about the orientational order of individual functional groups of molecules, and bout specific intra- and intermolecular interactions of these groups. The average IR absorbance's, $A_i$, of the particular vibrational modes, are determined by how the electric dipole moment of the system changes with the atomic oscillations. To the lowest order, the required quantities are proportional to the derivatives of the dipole moment with respect to the vibrational normal modes, $i$, of the system, evaluated at the equilibrium geometry. The IR absorbance of the $i_{th}$ vibrational mode is given by [31]:

$$A_i = \int_{v1}^{v2} A(v)dv = \frac{N\pi}{3c}\left[\frac{d\mu_i}{dQ_i}\right]^2 \quad (1)$$

where: $N$ is the number of molecules per unit volume, $\mu$ is the molecule dipole moment, and $Q_i$ is the normal coordinate corresponding to the $i_{th}$ mode.

For a dimer we choose a molecular system as follows: long axis (bow string) as a z-axis, x-axis normal to the bent plane and y-axis (bow arrow), which is the same as in [20].

In the LC phases the absorbance component of the selected vibrational band are related to the orientation of corresponding transition dipole moments.

$$A_{X,Y}/A_0 = 1 + S(\tfrac{3}{2}\sin^2\theta - 1) + \tfrac{1}{2}D\sin^2\theta\cos 2\phi$$
$$A_Z/A_0 = 1 + S(2 - 3\sin^2\theta) - D\sin^2\theta\cos 2\phi \quad (2)$$

Where: $A_0=(A_X+A_Y+A_Z)/3$, $S$ and $D$, are: orientational order parameters, of the long axis and molecular biaxiality, respectively, in the uniaxial nematic phase.

$$S = S_{zz}^Z, \quad D = S_{xx}^Z - S_{yy}^Z. \tag{3}$$

Following the Saupe ordering matrix [32,33], the parameter $S$ is a measure of the increase in compatibility of the molecule long axis with a nematic director while $D$ describes the rotational biasing of the short molecular axis.

Thus IR absorbance components can, at least in principle, bring information about the ordering in LC phase, via analyzing the order parameters $S$ and $D$. In the nematic phase the order parameter, $S$, follows the Mayer – Saupe model well [33]. However, in the $N_{TB}$ phase, it has been demonstrated that temperature dependence of the $S$-order parameter distinctly reverses its monotonic trend of increase to decrease at the transition temperature, N to $N_{TB}$ phase [34,35]. This is due to the rearrangement of molecules in the $N_{TB}$ phase as a helical structure appears. The orientation of the director gradually deviates from the helix axis, which is macroscopically observed as a reducing of the order parameter similarly like in the SmC phase.

Using polarized IR spectroscopy we can directly deliver the temperature dependence of the absorbance components in the temperature range of nematic and the $N_{TB}$ phases. For a planar sample we measured two components of the absorbance: one along the optical axis, as $A_Z$, and a second perpendicular to optical axis as an $A_Y$ component. For homeotropicaly aligned samples we measured an average of two perpendicular components $A_\perp=(A_X+A_Y)/2$ As it can be predicted from the $S$ parameter, we can expect distinctly different behavior of the absorbance components for the bands that the transition dipole is either longitudinal or transversal with respect to the core axis.

We analyzed the temperature dependence of the absorbance component for several bands with transition dipoles parallel and perpendicular to the terphenyl core in the range of the nematic and the $N_{TB}$ phases. Figures showing the temperature dependencies of the absorbance of the analyzed bands for selected dimers can be found in the supplementary materials (Fig.3s - Fig.5s). Also, the absorbance components for the structurally related MTC5 monomer were analyzed as reference data. By combining the IR results for the homogenous planar and the homeotropic alignment of the sample we can obtain all ($A_X$, $A_Y$, $A_Z$) components of the IR intensities. The average intensity $A_0$ and related transition dipoles were analyzed for series DTC5Cn in the temperature range of the N and the $N_{TB}$ phases and the MTC5 monomer in the N phase. Several vibrational bands are selected to be analyzed in the mid FTIR range, which provide significant dichroism of the band. These are the phenyl stretching band ($\nu CC$) at wavenumbers: 1485, 1460 and 1406 cm$^{-1}$. The combinational band at 905 cm$^{-1}$ that can be assigned to the phenyl in plane stretching with symmetric C-F stretching ($\nu CC+\nu_s CF$), phenyl in plane deformation with asymmetric C-F stretching at 890 cm$^{-1}$ ($\beta CC +\nu_{as} CF$) and phenyl out of plane deformation with asymmetric C-F stretching at 800 cm$^{-1}$ ($\beta CC +\nu_{as} CF$).

In order to determine the orientation of the short molecular axis of the dimer, we used the bands characteristic of an alkyl linker, which were assigned based on simulated vibrational spectra. To calculate the order parameter for the alkyl linker, we used: asymmetric deformation of the methylene groups called twisting vibration ($\gamma_{as}CH_2 + \beta CH$ in plane of the terphenyl) at 1320 cm$^{-1}$ and bending vibration - scissoring (in phase, $\beta_s$ CH$_2$) at the 1510 cm$^{-1}$ wavenumber. To determine the orientation of the terminal alkyl chains, we used symmetrical and asymmetric CH stretching vibrations of the of the methylene groups at 2890 and 2950 cm$^{-1}$, respectively ($\nu_s CH$, $\nu_{as} CH$).

## Results and discussion

### The terphenyl orientational order

The most prominent IR peaks belong to phenyl C-H and C-F vibrations in the terphenyl core. We can distinguish the C-F bands at 905 cm$^{-1}$ ($\nu CC+\nu_s CF$), 890 cm$^{-1}$ ($\beta CC +\nu_{as}CF$), and 800 cm$^{-1}$ which involve mostly C-F bond of the central phenyl in the terphenyl core. They correspond to the longitudinal transition dipole and the transverse transition dipole, respectively. For calculation of the orientational order of the terphenyl core it is useful to introduce a local system of reference, shown in the Fig. 2, as follows: the z-axis along the para axis of the terphenyl group, the x-axis is normal to the bent plane and the y-axis is normal to the z-x plane. Using the eq. (2) we can directly calculate the order parameter $S$ for the para axis of the terphenyl core of the DTC5Cn dimers (see Fig.3). The $S$ parameter found for DTC5C9 is in excellent agreement with the data obtained by $^{13}$C NMR spectroscopy by [20], both approach $S=0.49$ at the maximum.

The temperature dependences of the obtained order parameters can be related to the corresponding behavior for the monomer MTC5 (also shown in the Figure 3). For an odd number of carbons in the linking group, the low energy all trans conformation results in a bending angle of the mesogenic terphenyl groups; however due to the flexibility of the central hydrocarbon other conformations of the linker chain need to be considered. Finally we need to convert the orientational order of terphenyl core to that of molecular (long) axis. In order to do so we consider the angle $\beta_2=\beta/2$, that the terphenyl core makes with the bow string axis of the dimer; or in other words the shortest line between the ends of the mesogenic groups. The bend angle $\beta$ is expected to be $\beta_2=35.4°$, (see Fig.1b), for an all *trans* conformation of the linking chain as found from the DFT calculations. In the higher temperature range, the population of other conformers has a significant share, which means that the average bend angle is effectively reduced.

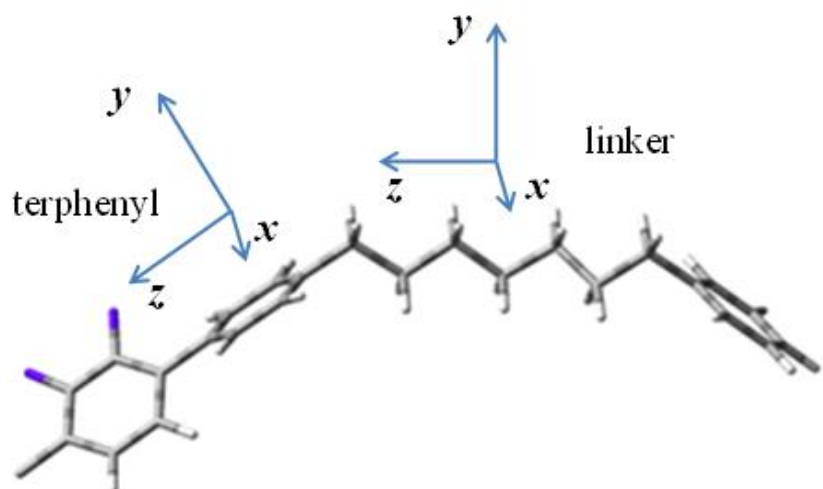

Fig.2. Frame of reference for molecular groups: terphenyl core and linking chain (only part of molecule is shown)

The corresponding second rank order parameter of the monomer scaled for a dimer, $S_{DIM}$, can be found as a product of the order parameter for the monomer, $S_{MON}$ by the Legendre polynomial $P_2(\cos\beta/2)$:

$$S_{DIM} = S_{MON}\ P_2(\cos\beta/2) \qquad (4)$$

Typically, it is expected that the orientational order is designated for the long axis of a molecule (bow string axis) of the dimer. We can convert the order parameter calculated for the terphenyl core only if we know the bending angle of the two mesogenic groups for a particular dimer. However, the angle is growing on decreasing temperature. For the short spaced dimers, DTC5C5 and DTC5C7 the order parameters are much lower ($\cong 25\%$) than that for the monomer. We found that it is not possible to scale $S_{MON}$ to $S_{DIM}$, by the Legendre polynomial $P_2(\cos\beta_2)$, as in eq (4), using one particular $\beta_2$ angle. In order to obtain the correct scaling it is necessary to increase the angle, gradually from 54° to 56° for the DTC5C5 and from 50° to 52° for the DTC5C7 on approaching the $N_{TB}$ phase.

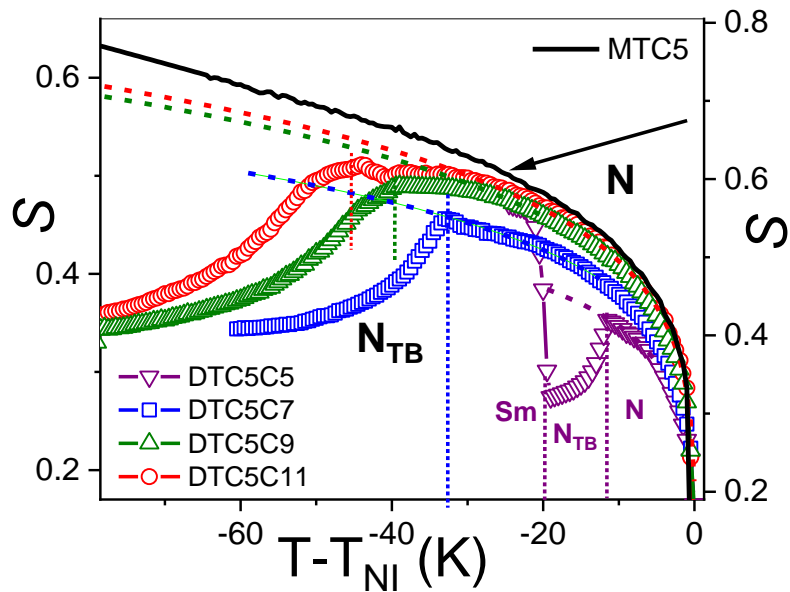

Fig.3. Comparison of S-order parameters for terphenyl core of DTC5Cn dimers with monomer MTC5 determined from 1485 cm$^{-1}$ band absorbance. *Dashed lines* – Fitting exp. data using the power law eq.(5). Symbols: ▽,□,△,○, are for n=5,7,9,11 respectively.

For dimers with a longer link DTC5C9 and DTC5C11, *S* parameters are initially closer to that of monomer, in the range of the N phase up to 25 K below the I-N transition temperature. This is probably due to a much higher flexibility of the linker. Scaling Legendre polynomials correspond to the bent angle 43.5° and 42° for DTC5C9 and DTC5C11, respectively. But on further cooling, the bent angles significantly increase up to 50°, for both dimers. Significantly, the observed temperature dependences of the angle $\beta_2$ are also similar for dimers DTC5C9 and DTC5C11. This leads to the conclusion that the population of more straightened conformers increases as the temperature rises in the nematic range [20]. When we compare the cases of successive homologues, it becomes clear that the bending angle $\beta_2$ is decreasing on increasing the number of carbons in the linker (*n*) for subsequent dimers. This is mostly the effect of the spacer flexibility increase as *n* increases. Ongoing from the $N_{TB}$ to the N phase, the molecules abandon the helix in favor of more straightened conformations [22], which leads to an easier formation of a uniaxial order and increases translational freedom. To achieve an overall straightened shape, the molecules must reduce the bend $\beta$ by twisting the spacer C–C bonds even further away from the trans state. For the above reason the temperature behavior of the order parameter cannot be correctly described by the Haller method [36]. The resulting critical exponent in that case may not have the proper physical significance, because it describes a system in which the share of components is variable.

For these reasons the temperature dependence of the nematic $S$-order parameter for the terphenyl moiety is flattened, and therefore the corresponding critical exponent, $\gamma$ is found significantly smaller than predicted by molecular models for monomers.

$$S = S_0 (1 - T/T_C)^{\gamma} \qquad (5)$$

One can fit eq. (5), a so-called power law, to the experimental data with the following critical exponents: 0.14, 0.16, 0.172, 0.175 for n=5,7,9,11 respectively. Fitted curves are shown by dashed lines in Fig.3. The approximation of the fitting curve can later be used as a reference for calculation of the tilt of the terphenyl core in the $N_{TB}$ phase.

**Orientational order of the central linker chain and molecular biaxiality of the dimer**

There are only few, rather weak IR-bands, which can be used for the calculation of the $S$ parameter of the aliphatic linking group. We have chosen the 1512 cm$^{-1}$ band, which is assigned to the CH$_2$ scissoring vibration (in phase, $\beta_s$ CH$_2$) and a rather complex band at 1320 cm$^{-1}$, which is assigned to the CH$_2$ twisting vibration mixed with the phenyl in plane deformation vibration ($\gamma_{as}$ CH$_2$ + $\beta$ CH). The transition dipole of the 1512 cm$^{-1}$ band is along the bisection of the CH$_2$ groups of the spacer chain, and this defines the y-axis of the molecular group (see Fig. 2). The transition dipole of the 1320 cm$^{-1}$ band is perpendicular to the long axis of the spacer, i.e. $\theta = 90°$ (in *all trans* conformation of the spacer) and its azimuthal angle is $\phi \cong 45°$. In this case the second term in eq. 2 is vanished (sin2$\phi$=0) and the $S$-parameter can be directly calculated from the 1320 cm$^{-1}$ band.

As the bent conformers are more dominant on the temperature lowering, the dimer linker becomes almost along the long axis of the dimer (bowstring axis). Therefore, the orientational ordering of the linking chain follows that of the overall long axis of the molecule. Figure 4 shows the experimental temperature dependence of the order parameter, $S$, for the dimer linker. Indeed, going from the isotropic to the nematic phase, the S parameters of the dimers coincide well with the S-parameter of the monomer, but gradually diverge from it on further cooling. This can be explained by the growing population of bend conformers, which diminishes the compatibility of dimers with the uniaxial director. Using this procedure, we directly obtained the quantitative value of the order parameter, (again we do not use Haller's method). In the range of nematic phase, the resulting order parameter $S$ is only 4-6% lower than obtained for the MTC5 monomer. Results for DTC5C9 are also in good agreement with data determined using diamagnetic anisotropy measurements [25] (parameter $S$ varies between 0.4 and 0.65) while the results obtained from optical birefringence [24] experiments give values as high as $S$=0.67. The latter high $S$ values, however, are obtained by applying the Haller method and surmise an orientational order higher than that found for the monomer. Fitting of the power law eq. (5) to the IR data the following critical exponents are obtained: 0.13, 0.16, 0.180, 0.184 for n=5,7,9,11 respectively, as shown by the dashed lines in Fig.4. These values are significantly lower than found for the monomer MTC5 ($\gamma=0.24$). The reason for that is that the number of bent conformers increases on reducing the temperature. The growing population of bending conformers reduces the $S$ parameter, see eq. (4) and flatten the temperature dependence of the order parameter for the dimer. Thus resulting critical exponent factor, $\gamma$, becomes significantly lower than expected based on the classical theoretical models of the nematic phase. This approximation of the fitting curves can be useful as a tool to calculate the tilt of the terphenyl core in the $N_{TB}$ phase. As can be seen in Figure 4 after the transition to the $N_{TB}$ phase, the upward trend of the order parameter is reversed.

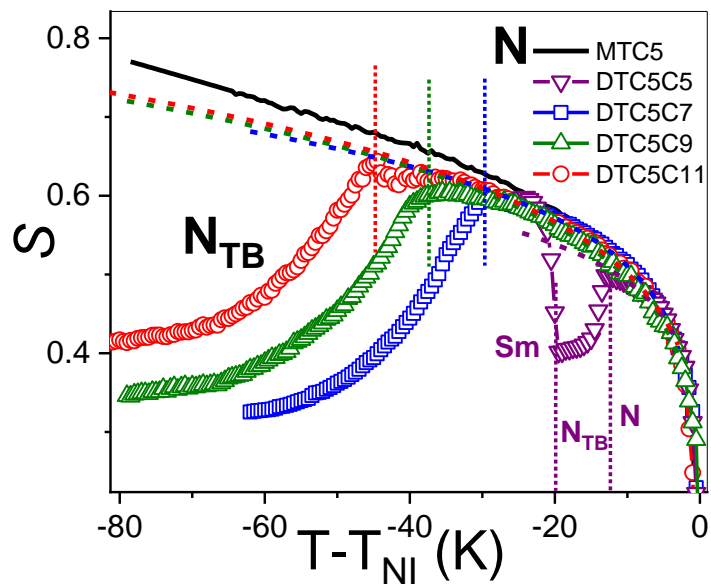

Fig.4. Temperature dependence of $S$-order parameters ($S = S_{zz}^Z$) of central linker for all DTC5Cn dimers, determined from 1320 cm$^{-1}$ band absorbance. *Black line* is $S$-order parameter of the MTC5 monomer as a reference. *Dashed lines* – Fitting the power law eq. (5) to the data. Symbols: ▽,□,△,○, are for n=5,7,9,11 respectively.

The corresponding order parameter in the $N_{TB}$ phase, $S_{TB}$ can be found in reference to that in N phase as a product:

$$S_{TB} = S_N P_2(\cos\theta_t) \qquad (6)$$

where: $\theta_t$ is cone angle of the helix.

Figure 5 shows the tilt angle of the central linker calculated using eq.(6), where $S_N$ is reproduced by eq. (5) using the fit parameters for the N phase extended over the $N_{TB}$ phase range. For shorter dimers (n=5,7) the tilt angle of the terphenyl core clearly vanishes on approaching the transition temperature, whereas for longer dimers (n=9,11) the transition region (10 K to 15 K above the $N_{TB}$ transition) can be distinguished, where the terphenyl core is already tilted. We note that the enthalpy of the transition to the $N_{TB}$ phase is a very weak for n=9,11 [22].

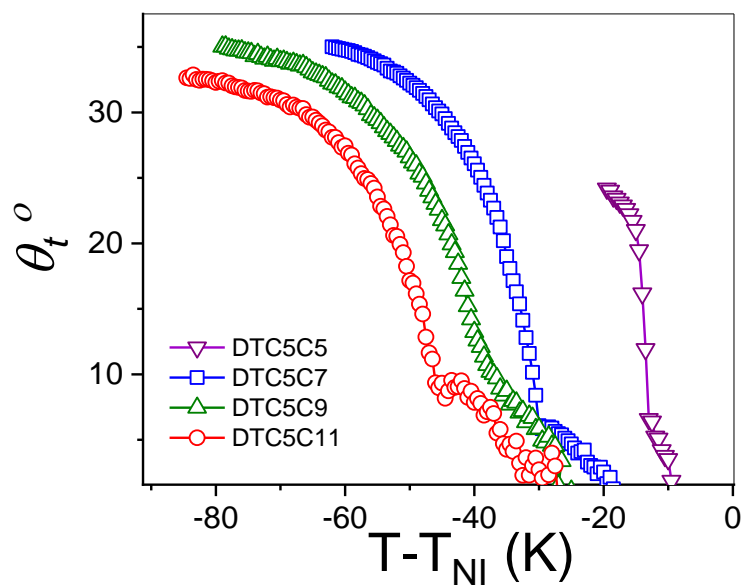

Fig.5. Tilt angle of the molecular long axis, determined from the absorbances of the 1320 cm$^{-1}$ band. Symbols: ▽,□,△,○, are for n=5,7,9,11 respectively.

In the next step, the absorbances of the two bands: 1512 cm$^{-1}$ and 1320 cm$^{-1}$ for the central linker were used to obtain the biaxiality of the short molecular axes. For both bands transition dipoles are normal to the long axis of the dimer, but they are at a different azimuthal angle, $\phi$. If both absorbances are combined, we are able to determine two elements of Saupe ordering matrix: $S_{xx}^Z$, and $S_{yy}^Z$. They describe how short axes of the central linker, $x$ and $y$, respectively, are oriented with respect to the $Z$-axis of the laboratory frame. The difference of the two parameters defines the molecular biaxiality, $D = S_{xx}^Z - S_{yy}^Z$, thus describing which one of two: $x$ and the $y$ axis are declined more from the laboratory $Z$-axis. Figure 6 shows the Saupe $S_{yy}^Z$ and $S_{bb}^Z$ and recalculated $S_{xx}^Z$ parameters for DTC5C7 dimer and also $S_{yy}^Z$ ($\cong S_{xx}^Z$) for the MTC5 monomer as a reference. In the nematic phase all $S_{ii}^Z$ ($i$=x,y,b) coincide with each other and also agree well with that of the monomer. In the N$_{TB}$ phase, however, both $S_{bb}^Z$ and $S_{yy}^Z$ decline from the trend of the monomer, because the central linker becomes tilted with respect to the $Z$-laboratory axis (helical axis). What is even more important, the $y$- axis remains preferably perpendicular to the $Z$-axis while the $x$-axis comes closer to the $Z$-axis. As a result the biaxiality parameter, $D$, which is a measure of such difference, $S_{xx}^Z - S_{yy}^Z$, is positive and furthermore it gradually increases on cooling in the N$_{TB}$ phase. Figure 7 shows the set of values for the molecular biaxiality parameter, $D$, for all the DTC5Cn dimers. In the range of the nematic phase, $D$ parameters are not significant, ($D\cong0.01$), with values similar to those observed for the MTC5 monomer, as the latter is assumed to show a uniaxial nematic phase. But on entering N$_{TB}$ phase $D$ for the dimers suddenly starts to grow on reducing the temperature. It is noted, that the values grow more steeply for the compounds with a shorter linker, as these have a larger bend angle, $\beta$, thus a higher shape biaxiality. A significant increase of the molecular biaxiality, $D$, is shown to be related to the growth of the twist-bend fluctuations of the director at the N-N$_{TB}$ transition temperature that induces the effective bend of the molecule and forms the helical structure.

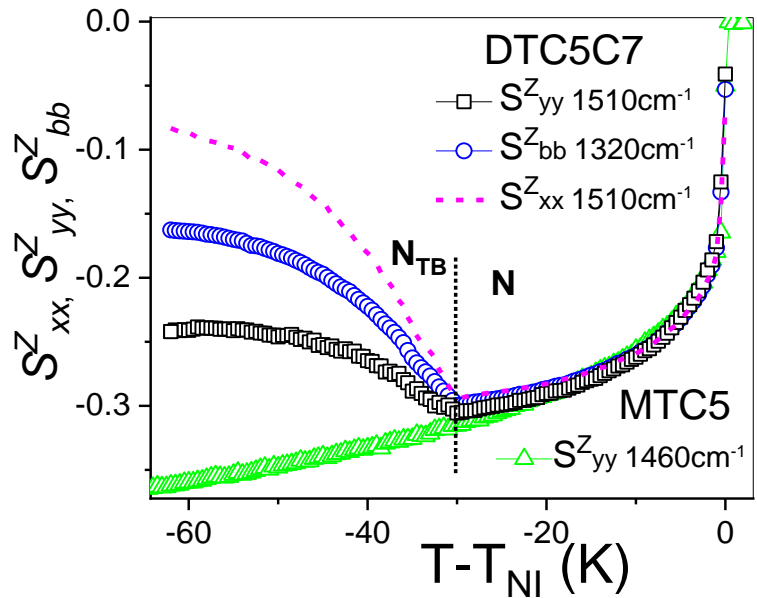

Fig.6. Saupe $S^Z_{yy}$ and $S^Z_{bb}$ and calculated $S^Z_{xx}$ parameters for the DTC5C7 dimer: □ -1510, ○- 1320, ---- calculated $S^Z_{xx}$ and also $S^Z_{yy}$ for the MTC5 monomer as reference- ▽.

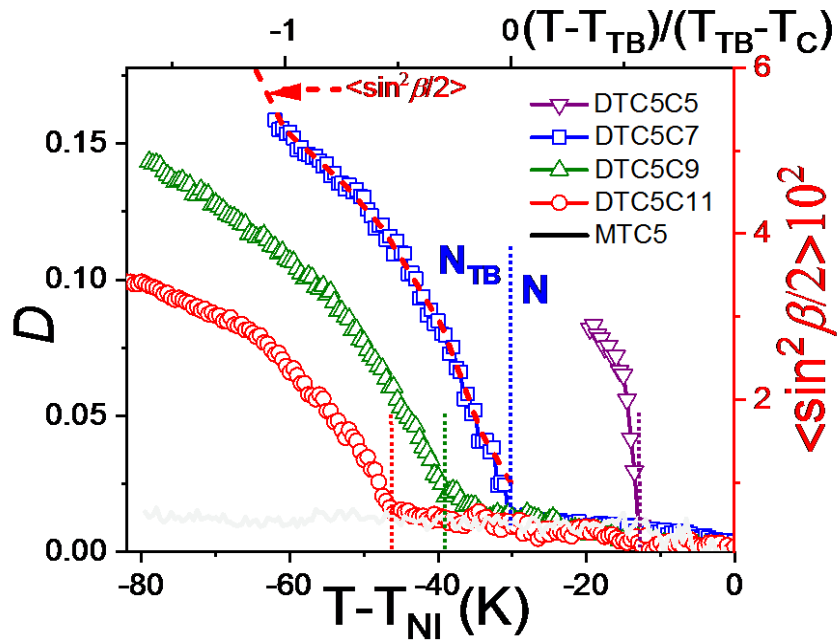

Fig.7. Molecular biaxiality parameter, $D$, solid lines: violet, blue, green, red for all DTC5Cn dimers, with n=5,7,9,11 respectively and for MTC5 monomer- solid black line. The local bending, $\langle \sin^2(\beta/2) \rangle$ calculated for the dimer mixture ('Se45')-[26] broken red line. (Bottom T-T$_{NI}$ and left Y axes are for DTC5Cn dimers but top T-T$_{TB}$/T$_{TB}$-T$_c$ and right Y axes are for the dimer mixture ('Se45').

Increasing $D$ on cooling indicates the hindering of the rotation of the short molecular axis. Such behavior is of a great importance for the phase behavior as biaxiality can be related to the local bending/curvature of the director that finally drives the transition to the nematic twist-bend phase.

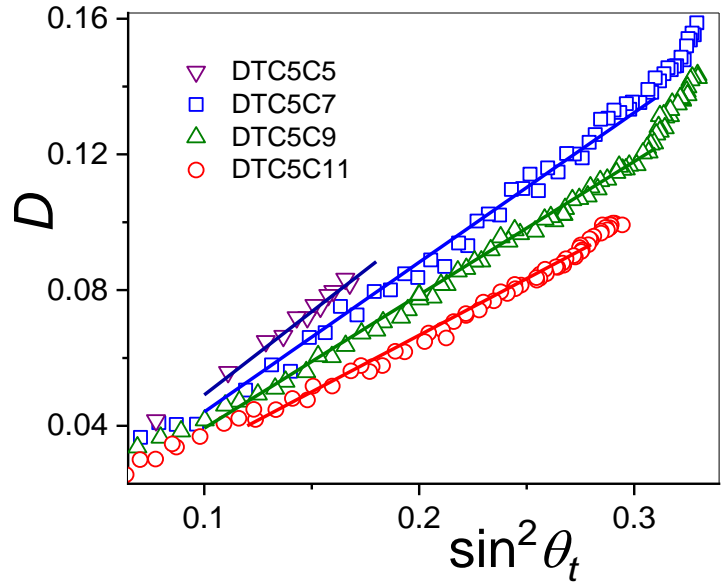

Fig.8. Molecular biaxiality parameter, $D$ vs. $\sin^2\theta_t$, for all DTC5Cn dimers: ▽,□,△,○, are for n=5,7,9,11 respectively and solid line are linear fits to the data (0,0 origin).

In order to analyze the result we can generally follow the model of geometry of the $T_{NTB}$ helix [19,22,26]. The proposed structure is mainly due to the twist of the biaxial molecular order around the long axes, quantitatively related to the curvature of the bend of the molecule. Nevertheless, we have to consider the fact that the molecules perform fast molecular rotations involving the short axes; spinning around the long molecular axis, precessions around the director or segmental spinning of the core ($>10^{-9}$ s$^{-1}$) also need to be taken into account. Therefore the bend of a molecule is locally seen as a statistical average over all orientations and possible conformational changes. In the uniaxial nematic phase the distribution averages the net bending to zero. But in $N_{TB}$ phase rotation becomes biased i.e. the biaxiality of the distribution of short axis appears, $D\neq 0$.

By definition of the pitch we have $q \equiv d\alpha/dz$, and from the geometry of the helix $\cos\theta_t = dz/ds$, we find $dz/ds = q\cos\theta_t$ [19] or $\alpha/d_2 = q\cos\theta_t$ [22], where $\alpha$ is the azimuthal rotation of the director per half the molecule length, $d_2$. Based on it, the bend magnitude of the helix is $B=q\cos\theta_t\sin\theta_t$ [19].

$$\sin(\beta/2) = tg(\alpha/2)\sin(\theta_t) \cong d_2 q \cos(\theta_t)\sin(\theta_t/2) \qquad (7)$$

It was already shown [19] that the local bending, $\sin(\beta/2)/(d_2\cos\theta_t)$, is linearly dependent on $\sin\theta_t$. In a molecular system, however, the bending vector has to be considered statistically, so its magnitude, $<\sin^2(\beta/2)>$, is a result of transforming the bend of the molecule $\sin(\beta_M/2)$ to the statistically averaged value:

$$\langle \sin^2(\beta/2) \rangle \cong D \sin^2(\beta_M/2) \tag{8}$$

As a result the molecular biaxiality parameter, $D$, due to eq (7) and eq (8) can be accordingly dependent on $\sin^2\theta_t$.

$$D \cong \langle \sin^2(\beta/2) \rangle / \left(d_2^2 \sin^2(\beta_M/2)\right) \cong C \sin^2\theta_t \tag{9}$$

where: $C=(d_2 \sin\beta/2)^{-2}$ is a slope of the "$D$ vs. $\sin^2\theta_t$" dependence shown in the Fig 9b.

Following eq. (7), it is possible to calculate the local bending $\langle\sin^2(\beta/2)\rangle$ for DTC5C7 using the data for the dimer mixture ('Se45') consisting of 55% DTC5C7 and 45% of a structurally related selenoether [26]. The result, which is shown as a $\langle\sin^2(\beta/2)\rangle$ by the red broken line in Fig.8., indicates a good coincidence with the biaxiality parameter, $D$, for the dimer DTC5C7. The molecular bending that can be calculated from eq. (7), $\beta_M/2 = 34°$, corresponds surprisingly well to the bending of the ground state conformer of DTC5C7, $\beta_M/2 = 35.4°$. In addition, if we analyses results for all dimers, it can be clearly concluded that the dependence "$D$ vs. $\sin^2\theta_t$" is becoming steeper for shorter linkers and less steep for longer linkers. In fact, the shortest dimer has highest aspect ratio (b/c) and furthermore elongation of the linker significantly reduces the molecular shape anisotropy.

The final conclusion is that the local bending $\langle\sin^2\beta\rangle$ and the magnitude of the director bend in the helix are clearly dependent on the molecular biaxiality and the molecular tilt. Figure 9 shows that the local director deformation is fully determined by the molecular biaxiality parameter ($D$).

We note that both perpendicular components of the absorbance, $Ax$, $Ay$ are very similar in the range of the nematic phase, except in the interval a few (dozen) degrees above the transition from the N to the $N_{TB}$ phase for DTC5C7 and DTC5C9 dimers. That was the reason to consider the nematic phase as uniaxial and also to simplify the structure analysis. Such an exception, however, can indicate the possibility that the biaxial nematic phase might appear in between the nematic uniaxial ($N_U$) and the twist-bend ($N_{TB}$) phase [37].

In fact, for proper analysis of the biaxiality of the phase we need to control the orientation of all spatial components, i.e. one parallel to the primary nematic director and two perpendicular, which are along secondary directors. Such conditions can be achieved through the anchoring effect or by an external electric field. This has been carried out previously for tetrapodes [38] and bent-core system [39].

**The orientational order of the terminal alkyl chain**

The terminal alkyl chains of dimers (end tails) are the most disordered part of the dimer in comparison to other segments of the molecule. This is primarily due to the fact that the chain (the *all trans* conformation) bends at an angle of 35° to the terphenyl core, and due to their high flexibility, they tend to be less ordered, as all possible conformations can coexist. For the calculation of the $S$ parameters we used the bands at the 2855 cm$^{-1}$ and 2930 cm$^{-1}$ wavenumbers, which are assigned to the C-H symmetrical and asymmetrical stretching vibration of the methylene groups ($v_sCH_2$, $v_{as}CH_2$), respectively.

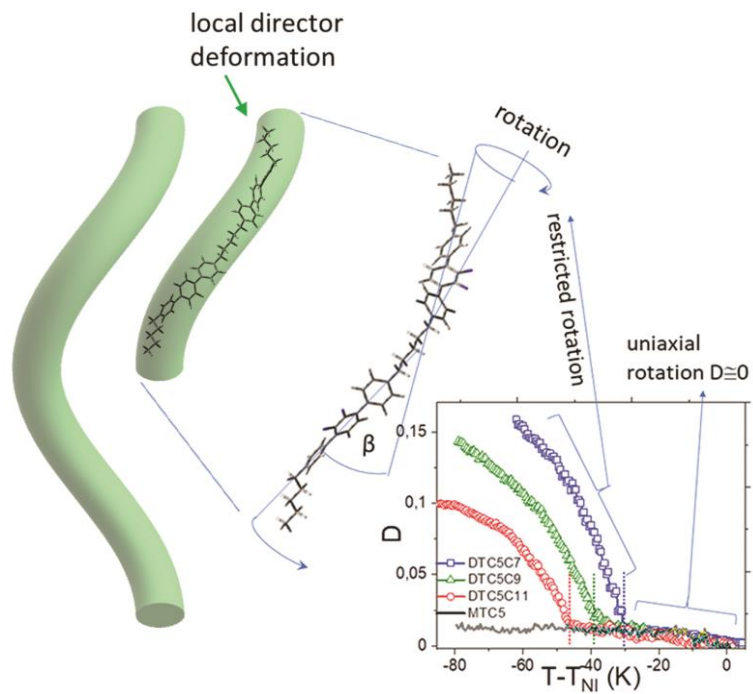

Fig.9. Schema presenting the local director deformation determined by the molecular biaxiality parameter D.

We define **y**-axis along the CH$_2$ bisection similarly as for the linker, as parallel to the symmetric transition dipole and **x**-axis perpendicular to the C-C-C plane of the alkyl chain. Unfortunately, there are both linker and tails contributing bands at the 2855 cm$^{-1}$ and 2930 cm$^{-1}$ and no reliable method to separate them can be found. Overall, the orientational order parameters *S* are found quite small, approaching 0.2 for n=7,9,11, but only 0.13 for n=5 (see Fig.10). This case is quite typical for order of the terminal chain in the N phase, so it can be considered that they describe the order of the terminal chains of the dimers. In the nematic phase the temperature dependences for n=7 and 9 are quite similar, contrary to n=11, where *S* parameter is significantly lower than former ones presumably due to higher flexibility of the longest linker. However suddenly, 5 K before entering N$_{TB}$ phase, it grows from 0.19 to 0.22, which probably indicates the process of nano-segregation (segmental correlation) of dimers before forming the helical structure. The efficient tilt angle of the tails in the N$_{TB}$ phase can be estimated using eq. (6). For n=9,11 they are 35° and 32° similarly like for the central linker, probably due to compatibility with helicoids shifted by $d_2$ along its axis [23] . For n=5,7 they are less than 25° in default of the compatibility.

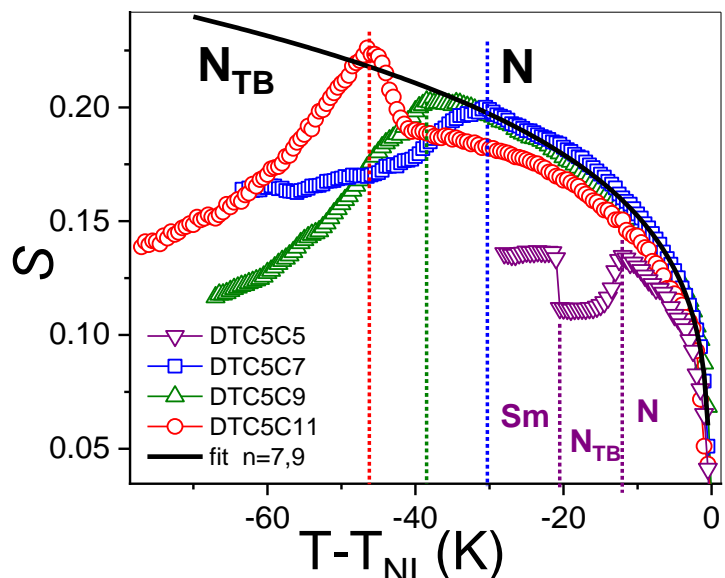

Fig.10. Temperature dependence of *S*-order parameters for the terminal tails for DTC5Cn dimers, determined from 2830 and 2950 cm$^{-1}$ band absorbances. Symbols: ▽,□,△,○, are for n=5,7,9,11 respectively and black solid line is a fit for n=7,9.

**Conclusion**

We used IR polarized spectroscopy to study orientational arrangements of the molecules in the nematic and the twist-bent phases for homologues series 2´,3´-difluoro-4,4´´-dipentyl-p-terphenyl dimers (DTC5Cn) and the corresponding monomer MTC5 as a reference. All spatial absorbance components ($A_X$, $A_Y$, $A_Z$) have been measured to obtain the information about the ordering in the N and the N$_{TB}$ phase. Several vibrational bands were chosen to analyses the orientation of different molecular groups of the dimers: the terphenyl core, central linker and terminal tails and also for the monomer as a reference. For all groups the orientational *S* and *D* ordering parameters have been determined, the former describing the ordering the long molecular axis while the latter restricted rotation (biasing) of the short axis. By comparing the temperature dependences of *S* parameters in the range of the nematic phase, we found that molecules remain in a bent conformation despite no measurable director bending. Bending is larger for a shorter linker ∼56° (n=5) but decreases when linker becomes longer ∼40° (n= 9,11). On increasing temperature the bending angle gradually decreases from ∼50° to ∼40° on approaching isotropic phase.

The biaxial order parameter, *D*, is found to be negligible in the N phase, then starts increasing on entering the N$_{TB}$ phase, following $\sin^2\theta$. The local director deformation, *B*, was found to be fully determined by the molecular biaxiality parameter *D*. The former vanishes at the transition to the N phase, similarly like *D*, because azimuthal rotation of the molecules is becoming isotropic. More importantly, the obtained parameters of the biaxial order, *D*, can predict the local director deformation, and consequently – the periodicity of the helical structure. Overall, using the geometric model and obtained molecular parameters, we can predict important properties of the N$_{TB}$ phase such as the cone angle, the bend of the director, and most importantly, the helical pitch.

## Conflicts of interest

There are no conflicts to declare.

## Acknowledgements

Authors (KM & AK) thank through the National Science Centre, Poland for Grant No. 2018/31/B/ST3/03609. All DFT calculations were carried out with the Gaussian09 program using the PL-Grid Infrastructure on the ZEUS and Prometheus cluster.